\documentclass[10pt, aps, prl, amssymb, amsmath, amsfonts, showpacs, floatfix, twocolumn, twoside, a4paper, superscriptaddress, longbibliography, nofootinbib]{revtex4-2}

\usepackage{mathrsfs}
\usepackage{siunitx}
\usepackage{comment}
\usepackage{soul}
\usepackage[T3,T1]{fontenc}
\DeclareMathAlphabet{\mathpzc}{OT1}{pzc}{m}{it}
\usepackage{mathrsfs}
\usepackage{bm}
\usepackage{microtype}
\usepackage{orcidlink}
\usepackage{multirow}
\usepackage{booktabs}
\usepackage{array}
\usepackage{tabularx}
\usepackage{braket}
\newcolumntype{C}{>{\centering\arraybackslash}X}

\usepackage{xcolor}
\usepackage{xspace}
\usepackage{nicefrac}
\usepackage{hyperref}

\hypersetup{
	colorlinks=true,
	citecolor=blue,
	linkcolor=red,
	urlcolor=blue,
}

\newcommand{\beq}{\begin{equation}}
\newcommand{\eeq}{\end{equation}}
\newcommand{\beqn}{\begin{eqnarray}}
\newcommand{\eeqn}{\end{eqnarray}}

\newcommand{\thetaJN}{\theta_{\rm{JN}}\xspace}
\newcommand{\fenv}{\ensuremath{f_{\rm env}}\xspace}

\newcommand{\GRAPPA}{\affiliation{Gravitation Astroparticle Physics Amsterdam (GRAPPA), University of Amsterdam, Science Park 904, 1098 XH, Amsterdam, The Netherlands}}
\newcommand{\Nikhef}{\affiliation{Nikhef, Science Park 105, 1098 XG Amsterdam, The Netherlands}}
\newcommand{\DESY}{\affiliation{Deutsches Elektronen-Synchrotron DESY, Platanenallee 6, 15738 Zeuthen, Germany}}
\newcommand{\Potsdam}{\affiliation{Institute of Physics and Astronomy, University of Potsdam, 28, Karl-Liebknecht-Straße 24/25, 14476 Potsdam, Germany}}
\newcommand{\DZA}{\affiliation{Deutsche Zentrum für Astrophysik DZA, Postplatz 1, 02826 Görlitz, Germany}}
\newcommand{\louvain}{\affiliation{Center for Cosmology, Particle Physics and Phenomenology - CP3, Université Catholique de Louvain, Louvain-La-Neuve, B-1348, Belgium}}
\newcommand{\belgium}{\affiliation{Royal Observatory of Belgium, Avenue Circulaire, 3, 1180 Uccle, Belgium}}

\begin{document}
\title{Biases from Astrophysical Environmental Effects in Standard-Siren Cosmology}

\author{Ioannis Liodis}\DESY \Potsdam \DZA
\author{Rodrigo Vicente}\GRAPPA
\author{Soumen Roy}\louvain \belgium
\author{Samaya Nissanke}\DESY \DZA \Potsdam \GRAPPA \Nikhef

\begin{abstract}
Next-generation ground based gravitational wave detectors will enable precision standard-siren cosmology, making systematics increasingly important. We investigate how line-of-sight acceleration, as may arise for compact binaries in dense astrophysical environments, biases measurements of the Hubble constant $H_0$. We simulate a catalogue of 50 binary neutron star bright sirens with known host redshifts, assuming the signals have been impacted by line-of-sight acceleration and analysing them with vacuum waveforms. Unmodelled line-of-sight acceleration biases the inferred luminosity distances and shifts the individual $H_0$ posteriors systematically towards lower values. Consequently, combining events narrows the posterior around a biased value rather than averaging away the systematic, even when the acceleration signs are assigned randomly. For our 50-event catalogues, affected fractions of approximately $10\%$ at $|a_\parallel|/c=10^{-6}\,\mathrm{s^{-1}}$, or $30\%$ at $|a_\parallel|/c=10^{-7}\,\mathrm{s^{-1}}$, produce offsets exceeding three posterior standard deviations in at least one quarter of the realisations. Considering larger catalogues further increases the significance of the effect. Our results indicate that environmental effects may constitute an important systematic for precision standard-siren cosmology and should be incorporated into future population analyses.
\end{abstract}

\maketitle

\textbf{\textit{Introduction.---}}
Measurements of the present expansion rate of the Universe reveal a discrepancy between early and late Universe inference (respectively, \emph{standard rulers}~\cite{Planck:2013pxb, BOSS:2013rlg, BOSS:2014hhw, Planck:2015fie, Planck:2018vyg, DESI:2024mwx, DESI:2025zgx} and \emph{standard candles}~\cite{Riess:2016jrr, Riess:2019cxk, Riess:2020fzl, Riess:2021jrx}) known as the Hubble tension. The Hubble tension may be due to systematics in the measurements of the different methods or due to new physics beyond the $\Lambda$CDM model. Gravitational wave (GW) observations offer an independent route to the problem, using compact binary mergers as \emph{standard sirens}~\cite{Schutz:1986gp, Holz:2005df}. 
GW signals directly encode the luminosity distance, $D_L$, without relying on the relative cosmic distance ladder. Redshift information can then be obtained from an identified electromagnetic (EM) counterpart and host galaxy, as for the binary neutron star (BNS) merger GW170817~\cite{LIGOScientific:2017vwq, LIGOScientific:2017zic, LIGOScientific:2017adf}, or inferred statistically from galaxy catalogs or source-population information~\cite{Schutz:1986gp, DelPozzo:2011vcw, Finke:2021aom, Gray:2021sew, Gair:2022zsa, Mastrogiovanni:2023zbw, Taylor:2011fs, Farr:2019twy, Mancarella:2021ecn}. 
With next-generation GW observatories, such as the Einstein Telescope (ET)~\cite{Punturo:2010zz, ET:2019dnz, ET:2025xjr} and Cosmic Explorer (CE)~\cite{Reitze:2019iox, Evans:2021gyd}, standard sirens are expected to reach sub-percent-level precision on the Hubble constant, $H_0$~\cite{Palmese:2025zku}. Their accuracy will depend critically on understanding and controlling systematics.

Astrophysical systematics can enter standard-siren inference in several ways. Peculiar velocities of host galaxies affect the inferred cosmological redshift, especially for nearby sources~\cite{Howlett:2019mdh, Nicolaou:2019cip, Mukherjee:2019qmm, Blake:2025etn}; weak lensing modifies the observed GW amplitude~\cite{Holz:2004xx, Mpetha:2024xiu}; the distance--inclination degeneracy makes bright-siren inference sensitive to viewing angle information, counterpart modelling, and selection effects~\cite{Chen:2020dyt, Chen:2023dgw}; dark sirens are sensitive to catalogue incompleteness and incorrect host galaxy weighting~\cite{Gray:2019ksv, Hanselman:2024hqy, Borghi:2025pav}; population methods depend on the assumed source distribution and its redshift evolution~\cite{Mastrogiovanni:2023zbw, Pierra:2023deu}. However there is yet another source of systematics, which remains less explored: waveform effects caused by dense local astrophysical environments.

Standard-siren analyses generally assume that vacuum waveform models recover an unbiased luminosity distance. This assumption may fail for binaries assembled in dense dynamical environments. Active galactic nucleus (AGN) disks provide a promising channel for the formation, hardening, and merger of black holes and neutron stars~\cite{Cheng:1999jr, McKernan:2012rf, McKernan:2014oxa, Bartos:2016dgn, Stone:2016wzz, McKernan:2020lgr, Zhu:2020wrd}. Drag from gas surrounding the binary, motion in the potential of the central supermassive black hole (SMBH), and interactions with nearby bodies can modify the GW signal~\cite{Meiron:2016ipr, Cardoso:2019rou, Sberna:2022qbn, Bonvin:2022mkw, CanevaSantoro:2023aol, Cusin:2024git, DuttaRoy:2025gnu, Tagawa:2025tfd, Santos:2025ass}. If omitted from the waveform model, these effects can be partially absorbed into the source parameters, leading to apparently precise but systematically biased measurements~\cite{Chen:2019jde, Chen:2020lpq, Roy:2024rhe, DeLuca:2025bph}. Line-of-sight acceleration (LoSA), generated by the central SMBH or by a nearby object, is expected to be among the dominant environmental imprints on stellar-mass compact binary signals in AGN disks~\cite{Zwick:2025wkt, Tagawa:2025tfd} and to be within reach of ET/CE~\cite{Vijaykumar:2023tjg, Lazarow:2024gdn} and Laser Interferometer Space Antenna (LISA)~\cite{Tamanini:2019usx}. 

A previous forecast showed that LoSA may bias the inferred $D_L$, and thus $H_0$, for \emph{individual} LISA sources~\cite{Tamanini:2019usx}; here, we quantify for the first time its \emph{population}-level impact for standard-siren $H_0$ inference with next-generation detectors. We inject BNS signals with LoSA into ET--CE networks and recover them with vacuum waveforms. We treat the sources as bright sirens with known host redshifts to isolate the effect of waveform systematics. For both acceleration signs, we find that the vacuum analysis systematically overestimates $D_L$ and so underestimates $H_0$. These shifts do not cancel when events are combined: the population posterior can narrow around a biased value even when acceleration signs are assigned randomly. Our results demonstrate that environmental waveform effects can limit the accuracy of standard-siren cosmology even when their impact is too small to be noticed in individual-event posteriors.\\

\textbf{\textit{Environmental effects on compact-binary mergers from AGN-channel.---}}
GW signals may be modified by kinematic effects associated with the motion of the binary's centre of mass, gaseous drag altering binary dynamics, and propagation through the gravitational field of the SMBH. These local effects are expected to result in very subtle changes in the waveform. Many of these are degenerate with the source parameters; for instance, a line-of-sight velocity is degenerate with a rescaling of the detector-frame masses and luminosity distance~\cite{Cusin:2024git}, while a transverse velocity can be absorbed by a change in binary orientation~\cite{Bonvin:2022mkw, Cusin:2024git}. Such uncorrelated effects across a population result mainly in a small broadening of the $H_0$ posterior. By contrast, time-dependent effects, such as LoSA and gaseous drag, introduce frequency-dependent deformations in the waveform that cannot easily be absorbed in the source parameters. The resulting loss in signal-to-noise ratio (SNR) in vacuum-waveform analyses may result in a systematic underestimation of the GW amplitude, and thus an overstimation of $D_L$.

To study how time-dependent environmental effects impact the $H_0$ inference, we consider the effect of LoSA (as argued in End Matter, we expect the effect from gaseous drag to be weaker). Let $\Psi_{\ell m}$ denote the phase of the $(\ell,m)$ mode of the frequency-domain waveform. If the LoSA, $a_\parallel$, varies negligibly over the observation, its leading contribution within the stationary-phase approximation (SPA) to the GW phase is~\cite{Roy:2026mco}
\begin{equation}\label{eq:LoSA_dephase}
    \Delta\Psi_{\ell m}^{\rm LoSA}(f)
    \approx
    -\pi\frac{a_\parallel}{c}\,f\,t_{\ell m}^2(f),
\end{equation}
where~$t_{\ell m}(f)\equiv t_{22}(2f/m)$ is the SPA stationary time associated with the $(\ell,m)$ mode, expressed in terms of that of the dominant $(2,2)$ mode\footnote{The amplitude of the frequency-domain waveform also receives subleading corrections~\cite{Roy:2026mco}, which we neglect here.}. This correction appears first at $-4$ Post-Newtonian (PN) order in an expansion of the phase in powers of velocity, $v/c = (\pi G M f/c^3)^{1/3}$~\cite{Tamanini:2019usx}. 

The LoSA distribution of stellar-mass compact binaries in AGNs is uncertain because it depends sensitively on the structure of the AGN disk. Ref.~\cite{Vaccaro:2025dzl} suggests that, for $M_{\rm SMBH}<10^8 M_\odot$, the majority of compact binaries in AGNs may form near migration traps. For a merger at a radius $R_{\rm trap}$, the characteristic acceleration is
\begin{equation}
    \frac{a_\parallel}{c}\sim 10^{-7}\, \mathrm{s^{-1}} \left( \frac{M_{\rm SMBH}}{10^6 M_\odot} \right)^{-1}\left( \frac{R_{\rm trap}}{10^3\,r_g} \right)^{-2}\,,
\end{equation}
with $r_g\equiv 2 G M_{\rm SMBH}/c^2$. Classical Type-I migration naturally produces traps at $R_{\rm trap}\sim10^3 r_g$, or smaller radii~\cite{Bellovary:2015ifg,Vaccaro:2025dzl}, while thermal torques typically lead to $R_{\rm trap}\sim10^5 r_g(10^6M_\odot/M_{\rm SMBH})$~\cite{Grishin:2023riv,Vaccaro:2025dzl}. Even in the latter case, Ref.~\cite{Vaccaro:2025dzl} finds that $\sim 10\%$ of compact binaries in AGNs may form at an inner trap at $R_{\rm trap}\sim10^2 r_g$ for $M_{\rm SMBH}\sim10^6M_\odot$. Binary--single interactions can produce larger accelerations, but these are expected to affect only a small fraction of the total AGN-channel merger rate~\cite{Tagawa:2025tfd}. Guided by these estimates, we adopt $10^{-9}\lesssim (|a_\parallel|/c)\,\mathrm{[s^{-1}]}\lesssim10^{-6}$ as the typical LoSA acting on compact-binary mergers in AGNs.\\

\textbf{\textit{BNS population.---}}
Forecasts show that $\sim 50$ bright sirens detected by Advanced LIGO--Virgo--KAGRA (LVK) could provide the precision required to address the Hubble tension \cite{Nissanke:2013fka, Chen:2017rfc, Feeney:2018mkj, Seto:2017swx, Chen:2024gdn}. ET--CE detector networks are expected to detect substantially larger samples within a single year~\cite{Chen:2024gdn}. In particular, $\sim 20$ loud BNS mergers at $z<0.1$ could enable an $ \sim 1\%$ measurement of $H_0$ with ET~\cite{Zhang:2020axa}.

We therefore adopt a conservative population of 50 bright sirens, selected from the one-year synthetic BNS catalog of Ref.~\cite{Gupta:2023lga} for a next-generation detector network. The selected events trace the underlying detected population. We further restrict the sample to $z\leq0.4$, for which ultraviolet, optical, and infrared electromagnetic counterparts are expected to be detectable~\cite{Sathyaprakash:2019rom}.

In particular, we started with the one-year BNS population of \cite{Gupta:2023lga} and used APR4\_EPP equation of state (EoS) from \texttt{LALSimulation} \citep{lalsuite}, to calculate the tidal parameters, instead of APR as in \cite{Gupta:2023lga}. Note that this EoS allows for a slightly lower NS maximum mass, requiring dropping $\sim3\%$ of the dataset. Note that the BNS population was generated assuming a local merger rate of 320 Gpc$^{-3}$ yr$^{-1}$. This value is high compared to recent estimations of 5.1–154.7 Gpc$^{-3}$ yr$^{-1}$ \cite{LIGOScientific:2026ctl}, but our further selections down to 50 events is not strongly bounded by it. Equivalently, assuming the lowest bound (5.1 Gpc$^{-3}$ yr$^{-1}$), we would need less than 5 months of observations to gather 50 BNS events with $z<0.4$.
\\

\textbf{\textit{Cosmological inference.---}}
We treat each binary as a bright standard siren and assume that its host redshift $z_j$ is known exactly from its EM counterpart~\cite{LIGOScientific:2017adf}. We adopt a spatially flat $\Lambda$CDM cosmology, neglect radiation, and fix $\Omega_m$ to its independently measured value (e.g. from baryon acoustic oscillations~\cite{DeLeo:2026svm}). The luminosity distance is, therefore,
\begin{equation}
D_L(H_0,z)=\frac{c(1+z)}{H_0} \int_0^z\frac{dz'} {\sqrt{\Omega_m(1+z')^3+1-\Omega_m}}\,,
\end{equation}
leaving $H_0$ as the only inferred cosmological parameter.

After marginalizing the GW likelihood over the remaining source parameters, the event-level posterior on $H_0$ and binary viewing angle, $\thetaJN$, is
\begin{align}\label{eq:single_posterior}
&p_j(H_0,\theta_{JN,j}\mid x_{{\rm GW},j},z_j) \nonumber \\
&\quad \propto \mathcal{L}_j\!\left[ D_L(H_0,z_j),\theta_{JN,j} \right] \pi(H_0)\pi(\theta_{JN,j}) \,,
\end{align}
where $\thetaJN$ coincides with the orbital inclination for the aligned-spin binaries considered here. For a catalogue of $N$ independent events, the combined posterior on $H_0$ is
\begin{equation}
p(H_0\mid\{x_{{\rm GW},j},z_j\}) \propto \pi(H_0)\prod_{j=1}^{N}\mathcal{L}_j(H_0)\,,
\end{equation}
with the marginalized likelihood
\begin{equation}\label{eq:L_j(H0)}
\mathcal{L}_j(H_0) \equiv \int d\theta_{JN,j}\, \mathcal{L}_j\!\left[ D_L(H_0,z_j),\theta_{JN,j} \right]\pi(\theta_{JN,j}) \,.
\end{equation}

\begin{figure*}[t]
    \centering
    \text{\texttt{CE + 2L ET}}\\
    \includegraphics[width=0.42\textwidth]{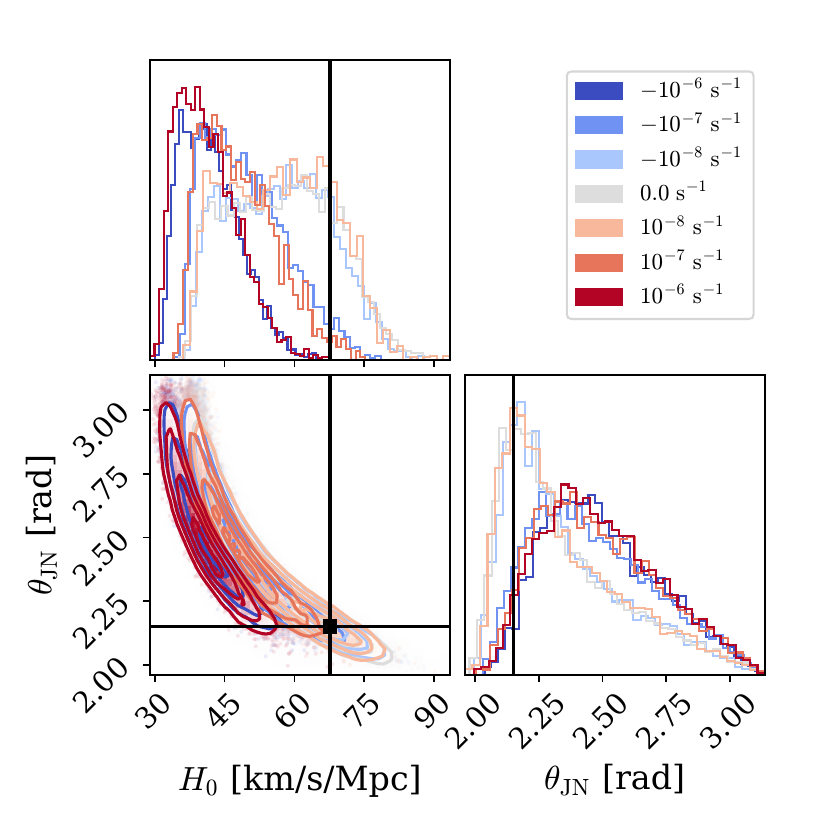}
    \includegraphics[width=0.42\textwidth]{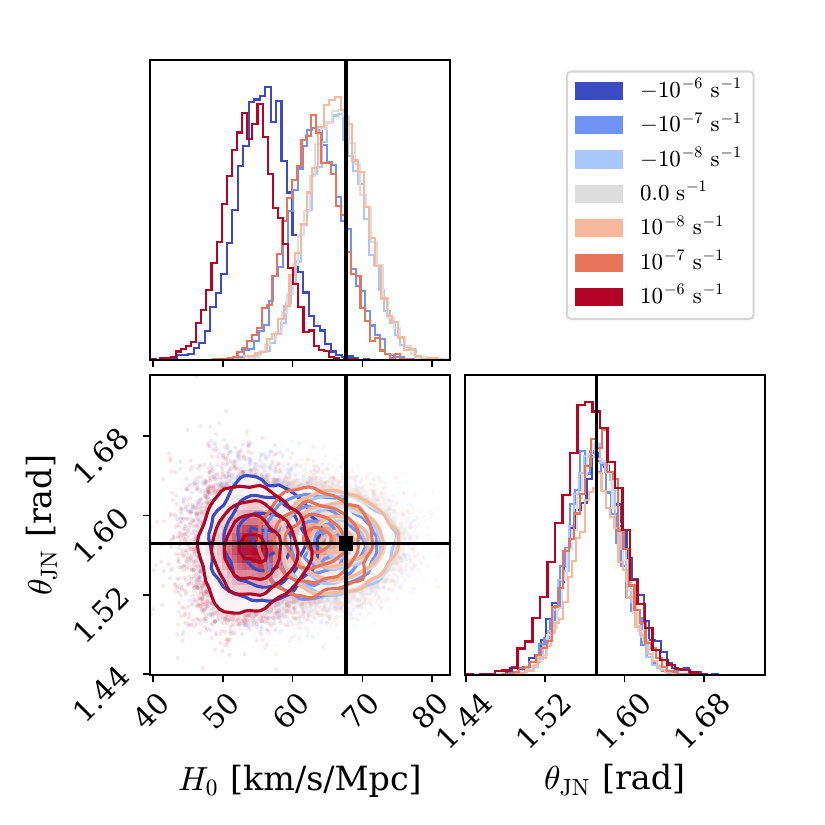}\\
    \text{\texttt{CE + $\triangle$ ET}}\\
    \includegraphics[width=0.42\textwidth]{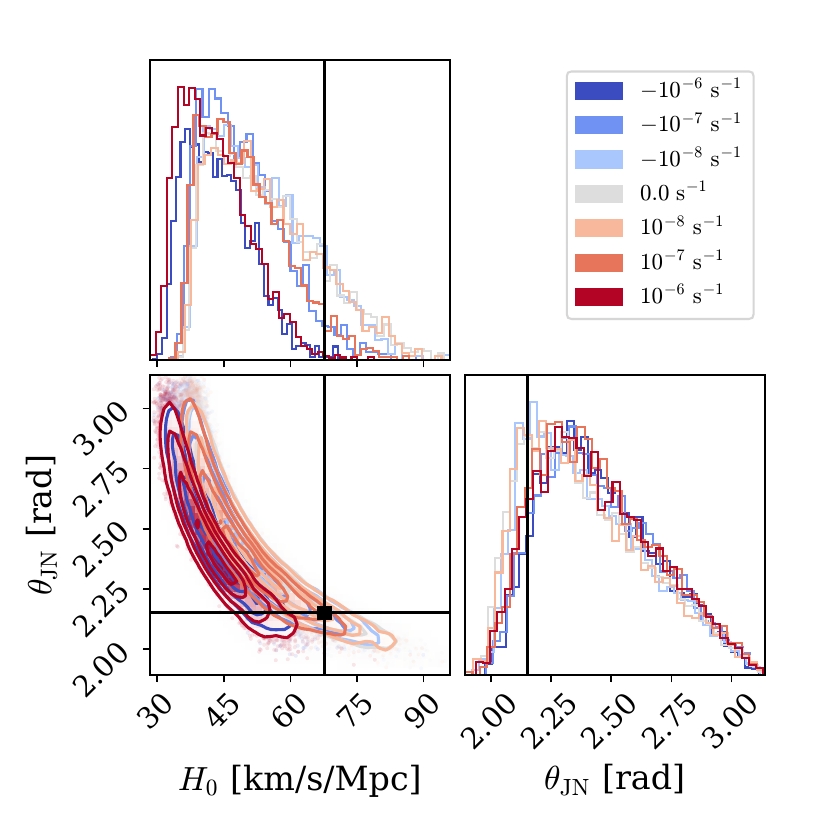}
    \includegraphics[width=0.42\textwidth]{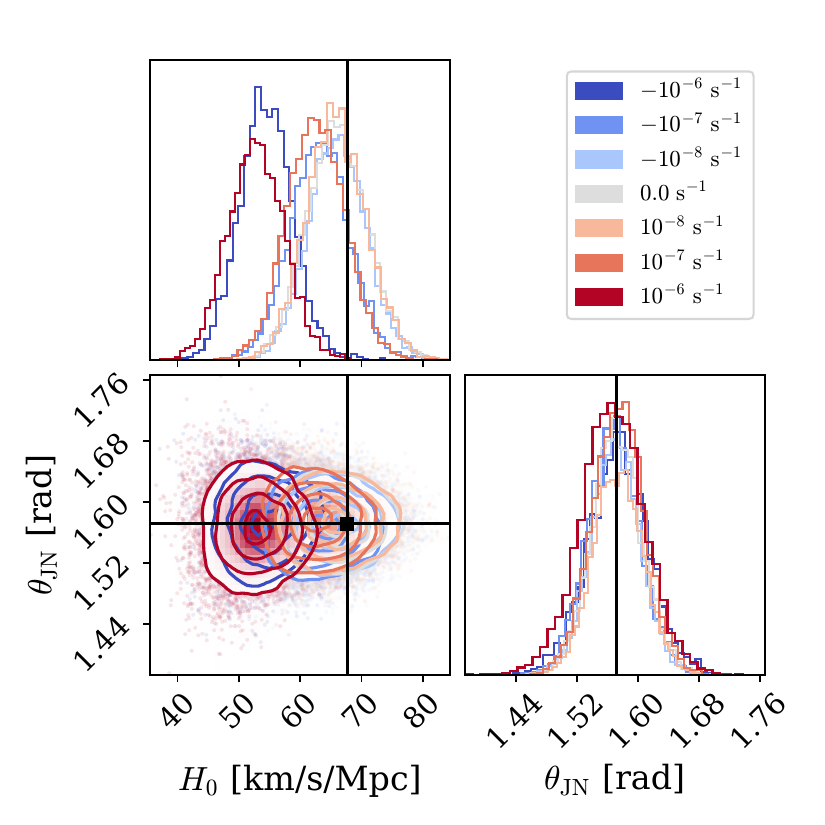}
    \caption{Joint and marginalised posteriors on $(H_0,\thetaJN)$ for two representative bright sirens: a closer to face-on event (left) and a nearly edge-on event (right). Signals are injected with six different LoSA values and recovered with a vacuum waveform, for \texttt{CE + 2L ET} (top) and \texttt{CE + $\triangle$ ET} (bottom). Blue and red curves denote negative and positive accelerations, respectively, with increasing colour intensity indicating larger $|a_{\parallel}|$; grey denotes the vacuum injection. Black lines and markers indicate the injected values.}
    \label{fig:bright_siren}
\end{figure*}

In practice, we reconstruct the likelihood $\mathcal{L}_j(H_0)$ by reweighting the $(D_L,\thetaJN)$ posterior samples obtained from parameter estimation. We map each distance sample as $D_{L,i}\mapsto H_{0,i}=H_0(D_{L,i},z_j)$ removing the parameter estimation prior $\pi_{\rm PE}(D_L)$.\footnote{The parameter estimation uses \texttt{Bilby}'s \href{https://bilby-dev.github.io/bilby/api/bilby.gw.prior.UniformSourceFrame.html}{\texttt{UniformSourceFrame}} distance prior.}\\

\textbf{\textit{Single-event analyses.---}} 
We perform zero-noise injections for six LoSA values characteristic of the AGN-channel, $a_{\parallel}/c\in\{\pm10^{-8},\pm10^{-7},\pm10^{-6}\}\,\mathrm{s}^{-1}$; we also injected signals in vacuum (for comparison). Signals were generated with \texttt{IMRPhenomXAS\_NRTidalv2} waveforms~\cite{Colleoni:2023ple} perturbed by LoSA [implemented through Eq.~\eqref{eq:LoSA_dephase}] and recovered with the corresponding vacuum model. Treating each source as a bright siren with known redshift then results in an independent posterior on $H_0$ [cf. Eq.~\eqref{eq:single_posterior}]. We consider two next generation detector networks: a) one $40\:\rm{km}$ CE and two 15km L-shaped ETs [\texttt{CE + 2L~ET}], and b) one 40km CE and one 10km triangle ET [\texttt{CE + $\triangle$ ET}] (see End Matter for details). We adopt a lower cutoff frequency of $5\:\rm{Hz}$ for the analysis.
Bayesian inference is performed with \texttt{Bilby}~\citep{Ashton:2018jfp,Romero-Shaw:2020owr} using the \texttt{Dynesty} nested sampler~\cite{Speagle:2019ivv} and the multibanding technique~\cite{Morisaki:2021ngj} for accelerated likelihood evaluations. We use 1000 live points with the \texttt{acceptance-walk} method and set \texttt{naccept}=60.

Figure~\ref{fig:bright_siren} shows the joint and marginalised $(H_0,\thetaJN)$ posteriors for two representative events of our population: one closer to face-on and another nearly edge-on. Events observed closer to face-on exhibit a strong correlation between $H_0$ and $\thetaJN$ (from the known distance--inclination degeneracy), which skews the posterior towards lower $H_0$ even for the vacuum injections~\cite{Nissanke:2009kt,Chassande-Mottin:2019nnz}. For $|a_{\parallel}|/c=10^{-8}\,\mathrm{s}^{-1}$, the individual posteriors remain nearly indistinguishable from the vacuum case. At $|a_{\parallel}|/c\geq10^{-7}\,\mathrm{s}^{-1}$, the vacuum recovery shifts the marginal $H_0$ posterior towards values below the injected value, irrespective of the acceleration sign, with the shift increasing in magnitude with $|a_{\parallel}|$. The viewing angle does not exhibit the same monotonic behaviour.\\

\begin{figure*}[t]
    \centering
    \text{\texttt{CE + 2L ET}}\\
    \includegraphics[width=0.99\textwidth]{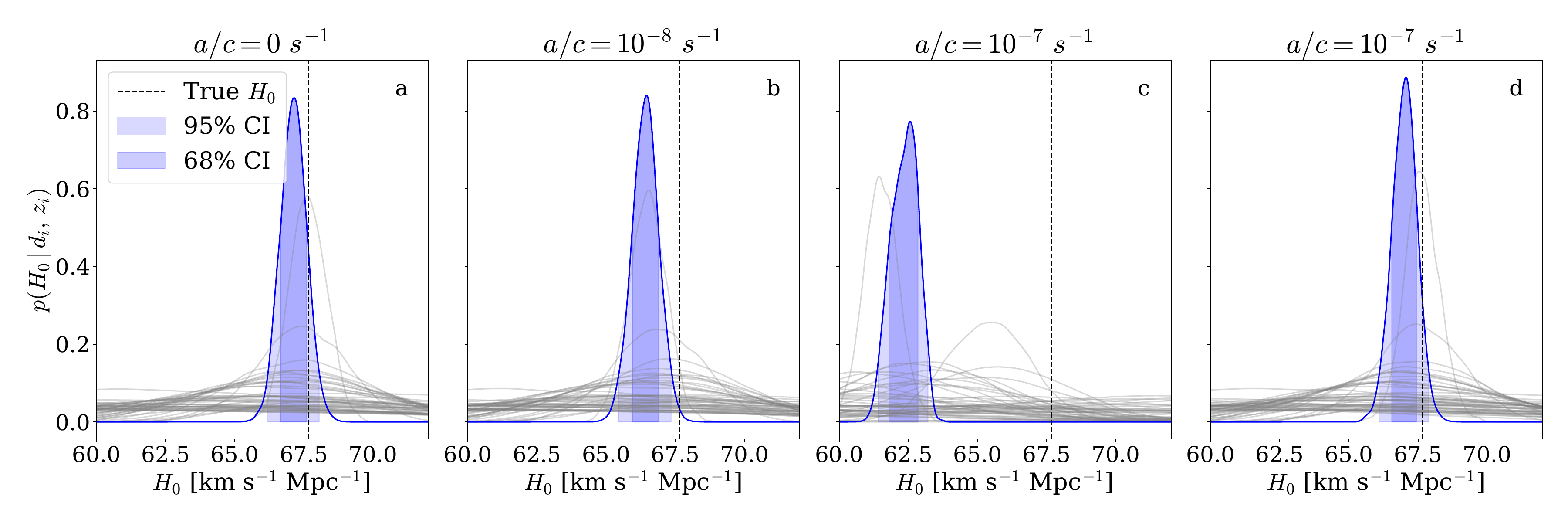}
    \text{\texttt{CE + $\triangle$ ET}}\\
    \includegraphics[width=0.99\textwidth]{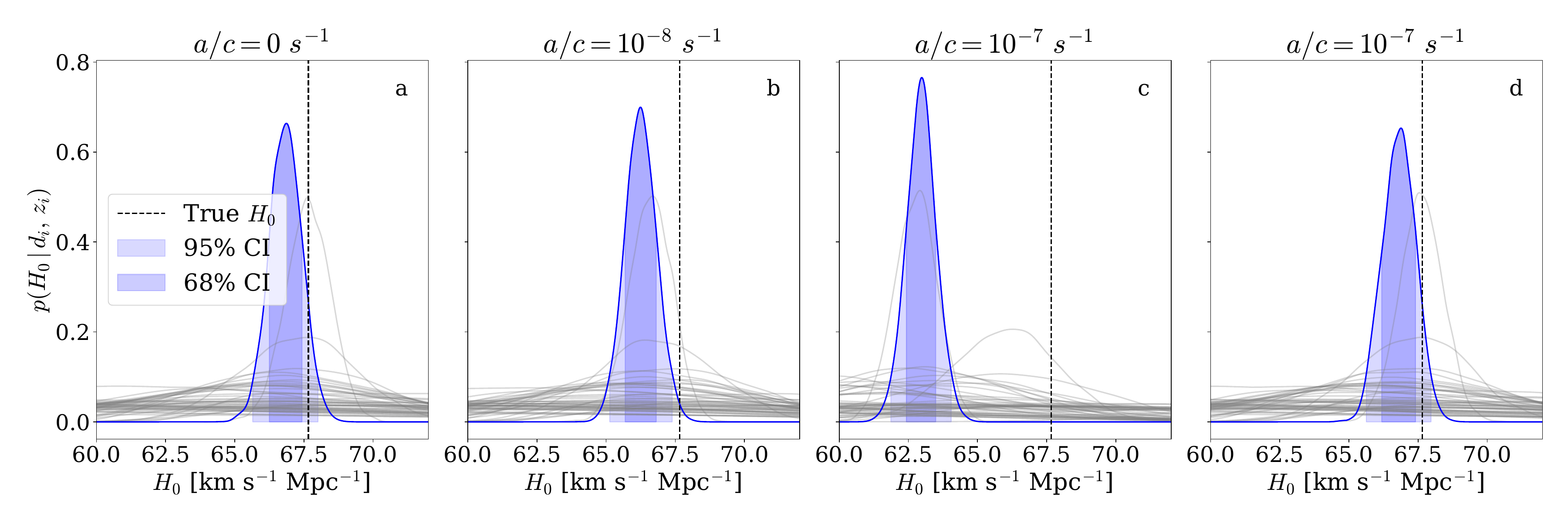}
    \caption{Event-level (thin grey curves) and combined (blue) $H_0$ posteriors for 50 bright sirens recovered with vacuum waveforms (panels a-c) and with the correct waveform (panel d). The signals are injected in vacuum (panel a) or with LoSA written in the title (panels b-d). The top row corresponds to \texttt{CE + 2L ET} detectors and the bottom row to \texttt{CE + $\triangle$ ET}. The vertical dashed line marks the injected $H_0$; dark and light shading denote the $68\%$ and $95\%$ credible intervals of the combined posterior.}
    \label{fig:combinedH0}
\end{figure*}

\textbf{\textit{Population analysis.---}} 
We perform the vacuum recovery for the full catalogue of $N=50$ bright sirens, considering $a_{\parallel}/c\in\{0,\pm10^{-8},\pm10^{-7},\pm10^{-6}\}\,\mathrm{s}^{-1}$. Figure~\ref{fig:combinedH0} compares the combined $H_0$ posterior for vacuum injections (panel a) with that obtained when every source has $a_{\parallel}/c=10^{-8}\,\mathrm{s}^{-1}$ (panel b) and $a_{\parallel}/c=10^{-7}\,\mathrm{s}^{-1}$ (panel c). Although the magnitude of the displacement varies across events, the individual posteriors shift predominantly towards lower $H_0$. Their product therefore narrows around a biased value rather than averaging back to the injected value: increasing the catalogue size improves the apparent precision without eliminating the waveform systematic. For comparison, in Fig.~\ref{fig:combinedH0} we also show the combined $H_0$ posterior with a LoSA-waveform analysis, for $a_{\parallel}/c=10^{-7}\,\mathrm{s}^{-1}$, using a uniform LoSA prior (panel d); as expected, the bias on $H_0$ is then resolved.

A realistic catalogue would contain a mixture of isolated and environmentally affected binaries, with accelerations drawn from an uncertain astrophysical distribution. We avoid identifying the AGN-channel fraction $f_{\rm AGN}$ with a specific LoSA model and instead introduce the fraction \fenv describing the affected fraction within the assumed bright-siren sample. For each benchmark magnitude $a/c\in\{10^{-7},10^{-6}\}\,\mathrm{s}^{-1}$, a fraction \fenv of the catalogue is assigned $a_{\parallel}=\pm a$, with the sign chosen randomly, while the remaining fraction $1-f_{\rm env}$ is assigned zero acceleration. We generate 50 catalogue realisations in each \fenv bin and combine their event-level likelihoods.
For each realisation $r$, we quantify the displacement of the combined posterior by the normalised bias
\begin{equation}
\mathcal{B}_{H_0}^{(r)} \equiv \frac{|\mu_{H_0}^{(r)}-H_{0,\mathrm{inj}}|} {\sigma_{H_0}^{(r)}}\,,
\end{equation}
where $\mu_{H_0}^{(r)}$ and $\sigma_{H_0}^{(r)}$ are respectively the mean and standard deviation of the combined $H_0$ posterior. Thus, $\mathcal{B}_{H_0}$ measures the systematic displacement in units of the reported statistical uncertainty.

\begin{figure}[t]
    \centering
    \text{\texttt{CE + 2L ET}}\\
    \includegraphics[width=0.49\textwidth]{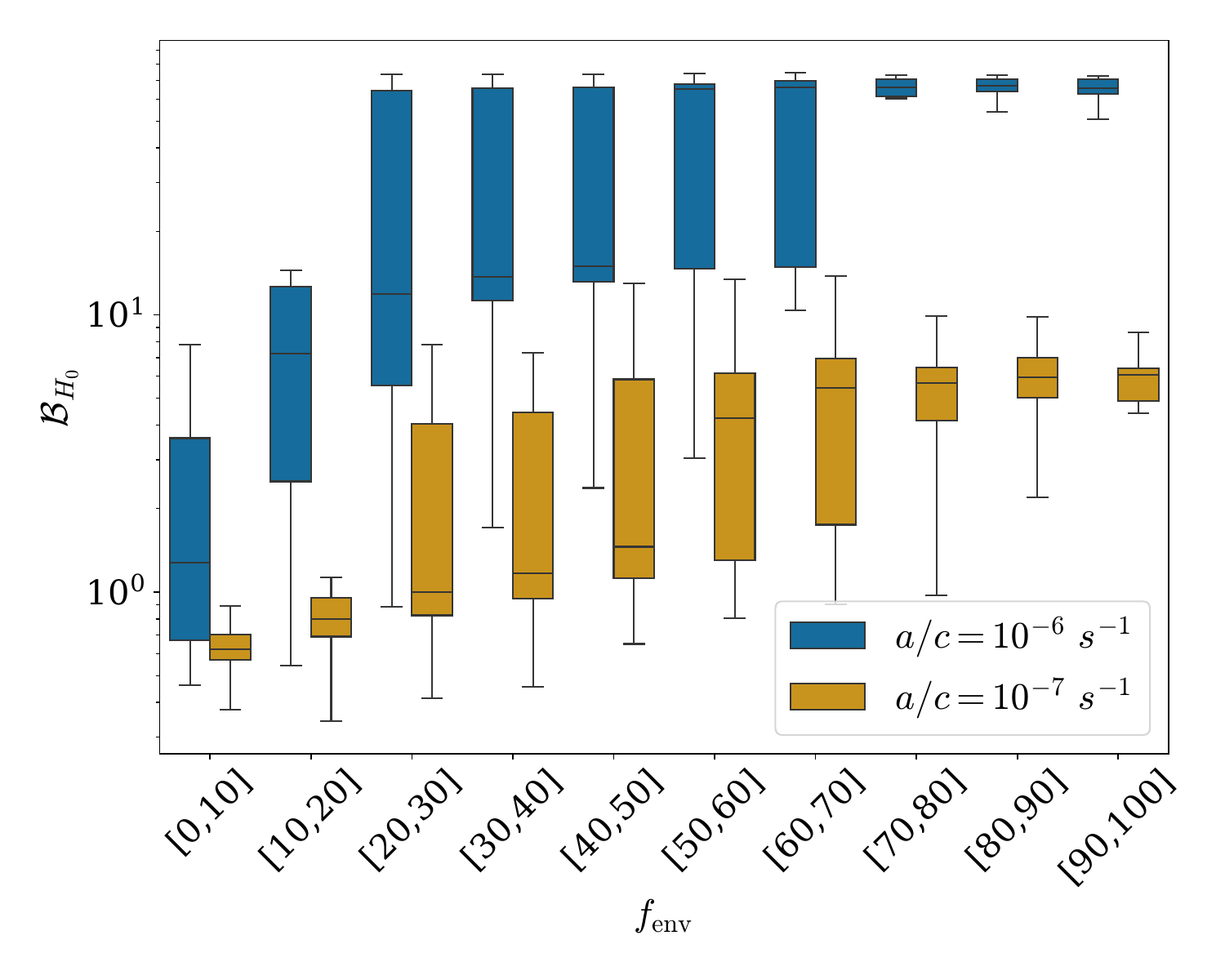}\\
    \text{\texttt{CE + $\triangle$ ET}}\\
    \includegraphics[width=0.49\textwidth]{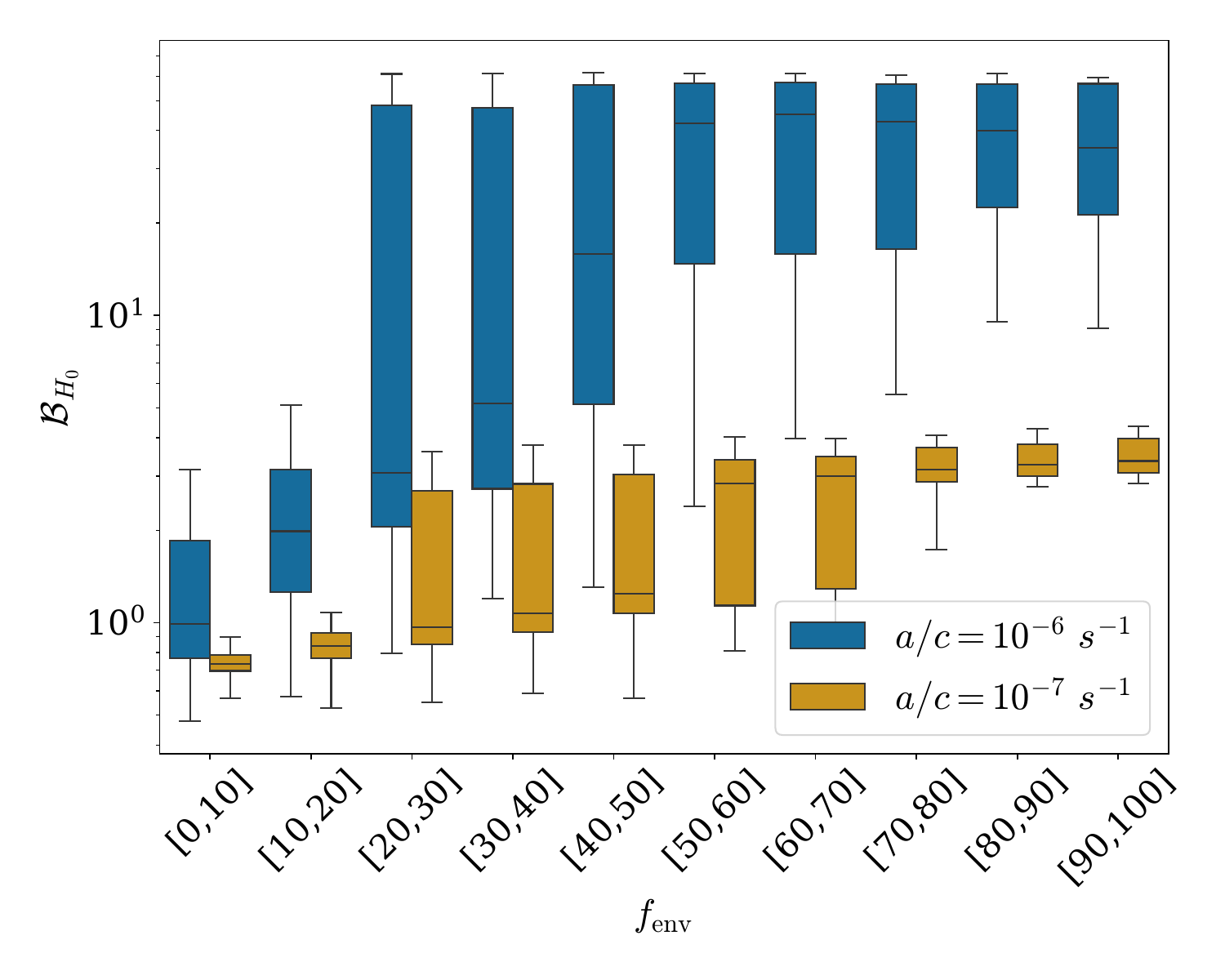}
    \caption{Bias in $H_0$ as function of the fraction of events affected by LoSA, \fenv, in a bright-siren population analysis, for \texttt{CE + 2L ET} (top panel) and \texttt{CE + $\triangle$ ET} (bottom panel). Boxes summarise $\mathcal{B}_{H_0}=|\mu_{H_0}-H_{0,\mathrm{inj}}|/\sigma_{H_0}$ across 50 catalogue realisations per \fenv bin for $|a_{\parallel}|/c=10^{-6}\,\mathrm{s}^{-1}$ (blue) and $10^{-7}\,\mathrm{s}^{-1}$ (orange), with acceleration signs assigned randomly. Central lines mark medians, boxes span the 25th--75th percentiles, and whiskers extend to the most extreme values within $1.5$ IQR.}
    \label{fig:H0bias_fenv}
\end{figure}

Figure~\ref{fig:H0bias_fenv} shows the distribution of the normalised $H_0$ bias across catalogue realisations as a function of \fenv. For both benchmark accelerations, the median $\mathcal{B}_{H_0}$ increases with \fenv and is markedly larger for the stronger LoSA. The broadest interquartile ranges occur at intermediate \fenv, where the bias depends strongly on whether the most informative sirens are environmentally affected. At large \fenv, these events are almost always affected and the distribution tightens around a large bias.

For $N=50$ bright sirens, the bias is substantial even when only $\sim10\%$ ($\sim30\%$) of the events are subjected to $|a_\parallel|/c=10^{-6}\,\mathrm{s}^{-1}$ ($10^{-7}\,\mathrm{s}^{-1}$): at these fractions, at least half of the realisations have $\mathcal{B}_{H_0}>1$, while at least one quarter have $\mathcal{B}_{H_0}>3$. As $N$ increases, the statistical uncertainties decrease approximately as $N^{-1/2}$ (for comparable independent events)~\cite{Nissanke:2009kt,Roy:2019phx}, whereas coherent systematic offsets need not average away; comparable bias significances may therefore arise at smaller \fenv or $|a_\parallel|/c$. Lowering the frequency cutoff from $5$ to $3\,\mathrm{Hz}$ similarly enhances the LoSA imprint.\\

\smallskip

\textbf{\textit{Discussion.---}} 
We have shown that LoSA, as may arise for compact binaries in AGN disks, can bias bright-siren measurements of $H_0$ even when acceleration signs are randomly distributed across the population. 
In our zero-noise injections, vacuum-waveform recovery systematically overestimates luminosity distances and shifts $H_0$ towards lower values.
We interpret this as a consequence of phase mismatch: vacuum templates cannot reproduce the frequency-dependent dephasing induced by LoSA, and the reduced signal overlap can favour a smaller recovered amplitude and hence a larger inferred distance. This interpretation suggests that similar biases may arise from other environmental effects that produce phase distortions not fully absorbed by changes in the source parameters.

We also demonstrated how combining events reduces the statistical uncertainty on $H_0$ without removing the systematic offset, only increasing its significance. This may cause even small environmental effects, which are not noticeable at the level of individual-event posteriors, to significantly impact standard-siren analyses. 
Including LoSA in the recovery model removes the bias in the configuration tested here. But beyond mitigating cosmological systematics, hierarchical analyses incorporating LoSA could constrain the fraction of accelerated mergers and their acceleration distribution. Combined with astrophysical modelling, these measurements could help constrain the contribution of the AGN channel and the existence and location of migration traps in AGN disks~\cite{Bellovary:2015ifg, Grishin:2023riv, Vaccaro:2025dzl}.

In summary, our results indicate that astrophysical environmental effects (such as LoSA) may constitute an important systematic for precision standard-siren cosmology, but also suggest an interesting opportunity for hierarchical Bayesian analyses. 
Several assumptions and opens questions suggest some possible research directions:
\begin{itemize}
    \item We adopt benchmark acceleration magnitudes and a common source population. Future analysis should include a distribution of accelerations and correlations between environment and source properties;
    \item We assume that each source has an identified counterpart and an exactly known host redshift and sky position. Counterpart selection and environmental biases in host identification remain to be assessed;
    \item We use quasi-circular waveforms. Degeneracies between eccentricity and LoSA~\cite{Roy:2026mco} may impact cosmological and environmental inference;
    \item The implications for dark sirens remain open: biases in distance, mass, and sky location may jointly affect cosmological inference.
\end{itemize}

\smallskip

\acknowledgments
\textbf{\textit{Acknowledgments}}
We thank the Helmholtz association for generous financial support.
The authors gratefully acknowledge the financial support provided by the German Federal Ministry of Research, Technology and Space (BMFTR) in the framework of the Knowledge creates perspectives for the region!, for the project StStG – DZA – Aufbauphase: Deutsches Zentrum für Astrophysik, Großforschungszentrum in der sächsischen Lausitz: Aufbauphase 2026, grant number 03WSP1746.
The authors gratefully acknowledge the computing time made available to them on the high-performance computers at the German Center for Astrophysics (DZA). Access to these computational resources and the associated technical support was essential for carrying out the simulations and obtaining the results presented in this work. The DZA is currently being established as part of a project funded by the Federal Ministry of Research, Technology and Space (FMRTS/BMFTR).
This scientific paper was supported by the Onassis Foundation - Scholarship ID: F ZU 042/2 2025-2026. 
IL would also like to thank the Lilian Voudouri Foundation for their master's scholarship, the Onassis Foundation for their master's and doctoral scholarship, as well as Deutsches Elektronen-Synchrotron (DESY).
RV gratefully acknowledges the support of the Dutch Research Council (NWO) through an Open Competition Domain Science-M grant, project number OCENW.M.21.375.

\bibliography{References.bib}

\section{End Matter}\label{sec:supplement}

\textbf{\textit{Effects from gas torques.---}}
The effects from gas in the dynamics of merging compact binaries in AGN disks cannot be fully captured by simple analytic models. Hydrodynamical simulations have explored this evolution primarily before GW emission dominates (e.g.,~\cite{Li:2022pnc, Li:2022eup, Li:2023gyv}). If the binary opens a cavity in the AGN disk and develops a circumbinary disk, the gas torque can be approximated by
\begin{equation}
    T_{\rm cbd} \approx G \xi \dot{M} M/v\,,
\end{equation}
where~$\dot M$ is the mass accretion rate and $\xi$ is a dimensionless torque coefficient, with simulations supporting values of $\xi\sim 0.1$.

Mass growth and the circumbinary disk torque both modify the orbital chirp, producing a leading correction to the phase of the dominant $(2,2)$ mode
\begin{equation}
    \Delta \Psi^{\rm gas} \approx -\frac{75}{851968}\frac{G \dot{M}}{c^3 \eta^2} \left(1-8\xi\right) \left(\frac{v}{c}\right)^{-13} \,,
\end{equation}
where $\eta$ is the symmetric mass-ratio.
Like LoSA, this correction enters at $-4$\,PN order in the waveform phase. For a nearly equal-mass binary ($\eta \simeq 1/4$), the ratio of their magnitudes is
\begin{align}
    \left|\frac{\Delta \Psi^{\rm gas}}{\Delta \Psi^{\rm LoSA}}\right| &\approx\frac{3}{13}\left(32 \xi-1 \right) \frac{\dot M/M}{|a_\parallel|/c}\,,\\
    &\sim 5 \times 10^{-9}\left(\frac{0.1}{\epsilon}\right)\left(\frac{10^{-7}\,\mathrm{s^{-1}}}{|a_\parallel|/c}\right)\,,
\end{align}
where the numerical estimate assumes $\xi=0.1$ and an Eddington accretion with radiative efficiency $\epsilon$. Under these assumptions, gas-induced dephasing is very much smaller than the LoSA contribution at the accelerations considered here.

\textbf{\textit{Detector networks.---}}
We consider two detector networks, each combining a 4\,km CE detector with either a 10\,km triangular ET [\texttt{CE + $\triangle$ ET}] or two 15\,km L-shaped ET detectors [\texttt{CE + 2L ET}]. The triangular ET is placed in the EMR region ($50.7600^\circ$ N, $5.9171^\circ$ E; elevation 160\,m), with $x$- and $y$-arm azimuths of $70.5674^\circ$ and $130.5674^\circ$, respectively. The L-shaped detectors are placed Lusatia ($51.323^\circ$ N, $14.2349^\circ$ E; elevation 100\,m) and Sardinia ($40.5167^\circ$ N, $9.4167^\circ$ E; elevation 187\,m), with respective $(x,y)$-arm azimuths of $(45^\circ,135^\circ)$ and $(0^\circ,90^\circ)$. For CE, we adopt the default geometry provided by \texttt{bilby.gw.detector.InterferometerList}.

\begin{figure}[ht]
    \centering
    \includegraphics[width=0.49\textwidth]{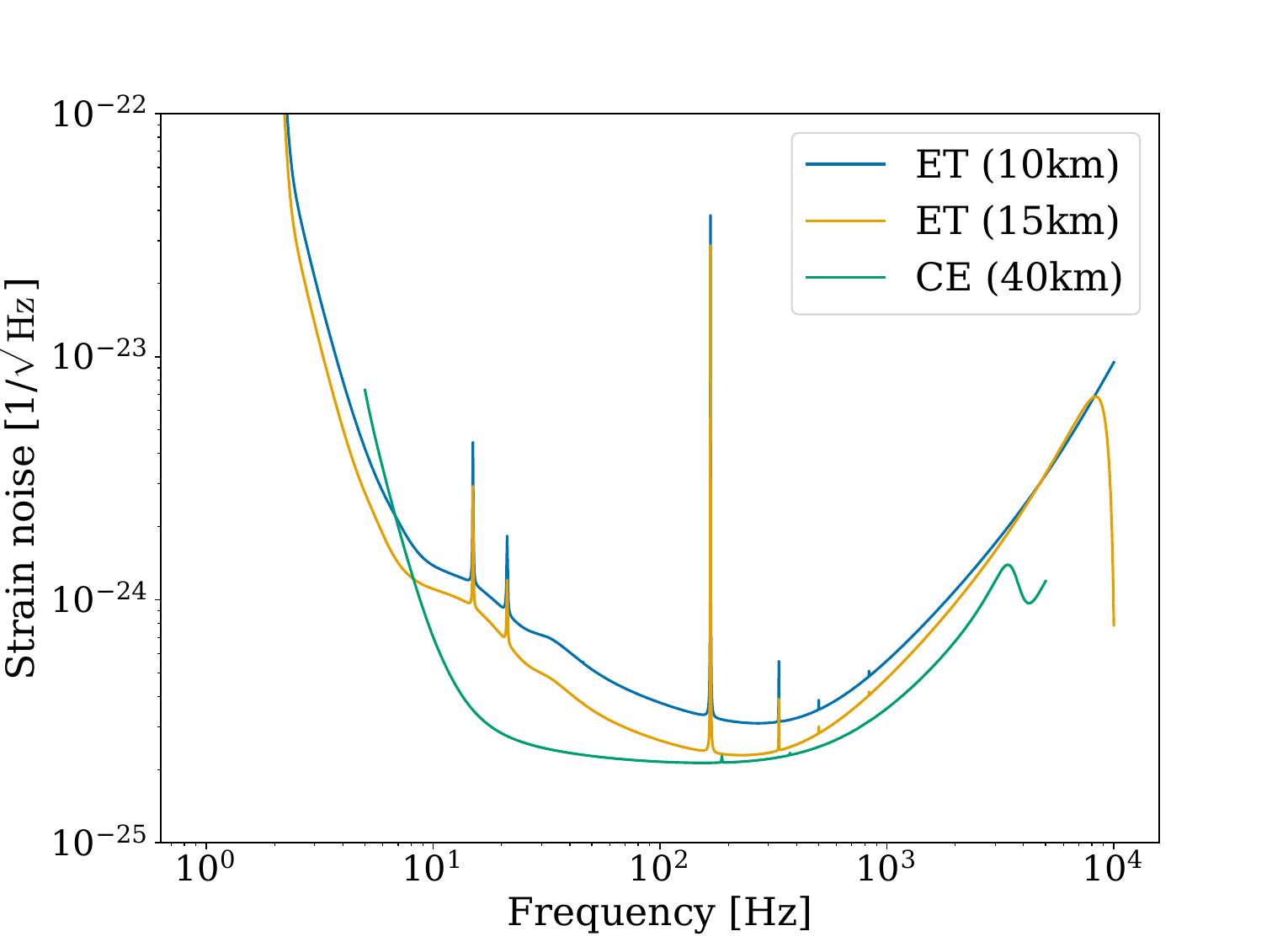}
    \caption{Sensitivity curves for the detectors considered in this work.}
    \label{fig:sensitivities}
\end{figure}

We used a lower frequency cutoff of 5\,Hz and the sensitivity curves shown in Fig.~\ref{fig:sensitivities}. Both ET 10\,km and 15\,km correspond to the xylophone configuration (high- and low-frequency)~\cite{ETSensitivity}.

\end{document}